\documentclass{JAC2003}

\usepackage{graphicx}
\usepackage{epsfig}

\newcommand{\bea}{\begin{eqnarray}} 
\newcommand{\eea}{\end{eqnarray}}

\begin{document}
\title{Hadronic cross section from radiative return\thanks{Presented
 by H. Czy\.z at
 Workshop on $e^+e^-$ in the 1-2 GeV range: Physics and Accelerator
 Prospects, 10-13 Sept. Alghero (SS), Italy.
 Work supported in part by TARI project HPRI-CT-1999-00088 and 
 Polish State Committee for Scientific Research
  (KBN) under contract 2 P03B 017 24.}}

\author{Henryk Czy\.z\thanks{czyz@us.edu.pl} 
 and Agnieszka Grzeli{\'n}ska\thanks{agrzel@us.edu.p}\\
 Institute of Physics, University of Silesia, PL-40007 Katowice, Poland.}

\maketitle

\begin{abstract}
  The impact of final-state radiation (FSR) on the radiative
 return method for the extraction of the $e^+e^-$ hadronic cross section
 is discussed in detail and experimental tests of the model
 dependence of FSR are proposed for the $\pi^+\pi^-$ hadronic final state.
\end{abstract}

\section{Highlights of the radiative return method}

  Knowledge of the cross section of the electron--positron
  annihilation into hadrons in the low energy region
   is crucial
 for predictions of the hadronic contributions to $a_\mu$, the anomalous 
 magnetic moment of the muon, and to the running of the electromagnetic
 coupling
(For reviews see 
 e.g.  [1-7];
%\cite{a_mu1,Jegetc.,Melnikov:2001uw,Davier:2002dy,HMNT02,Davier:2003,Nyffeler:2003};
 the most recent experimental result for $a_\mu$ 
  is presented in \cite{Bennet}).
 Measurements of the electron--positron hadronic cross section
 were traditionally performed  by varying the 
 beam energy of the collider, but the 
 \({\mathrm  \Phi}\)- and B-meson factories
 allow to use the radiative return to explore the whole energy
 region from threshold up to the energy of the collider. 
 Even if the photon radiation 
 from the initial state reduces the cross section by a factor
 ${\cal O}(\alpha/\pi)$, this is easily compensated by the 
 enormous luminosity of these `factories'. A number of experimental results
 based on the radiative return
 was already published [9-17]
%\cite{Aloisio:2001xq,Denig:2001ra,babar,Barbara:Morion,Berger:2002mg,Venanzoni:2002,Achim:radcor02,KLOE:2003,Blinov},
 and in the near future one can expect
 much more data covering large variety of hadronic final states.

 The radiative return method \cite{Binner:1999bt} (see also \cite{Zerwas}),
  relies on the following factorisation property of the cross section

\bea
 &&\kern-30pt\frac{d\sigma}{dQ^2d\Omega_\gamma}
 \left(e^+e^- \to \mathrm{hadrons} + \gamma\right) = \nonumber \\
&&H(Q^2,\Omega_\gamma) \ \sigma(e^+e^-\to \mathrm{hadrons,Q^2}) \ ,
 \label{rr}
\eea

\noindent
 where $Q^2$ is the invariant mass of the hadronic system, $\Omega_\gamma$ 
 denotes the photon polar and azimuthal angles, and the function
 $H(Q^2,\Omega_\gamma)$ is given by QED lepton-photon interactions,
 thus known in principle with any required precision.
 The formula (\ref{rr}) is valid for 
 a photon emitted from initial state leptons (ISR)
 and what is more important
 similar factorisation formula applies for the emission  
 of an arbitrary number
 of photons or even lepton pairs \cite{KKKS88,CN03}
  from initial state leptons. 
  If there is no FSR contribution,
 one can, by
 measuring the $Q^2$ differential cross section
 of the process 
 $e^+e^- \to \mathrm{hadrons} + \mathrm{photons} +
 (possibly)\ \mathrm{lepton \ pairs}$
  and knowing function 
 $H(Q^2,...)$, extract the value of $\sigma(e^+e^- \to \mathrm{hadrons})$.
 The $...$ in $H(Q^2,...)$ stand for the phase space variables of
 photons and/or lepton pairs. 
 In practice that cross section is measured within 
 a given experimental setup, corresponding to an integral  
 over a complicated phase space of photons and/or lepton pairs.
 As a result
 the use of  Monte Carlo event generators for extraction of 
 the $\sigma(e^+e^-\to \mathrm{hadrons,Q^2})$ become 
 indispensable. Such Monte Carlo programs 
 (EVA \cite{Binner:1999bt,Czyz:2000wh}, 
 PHOKHARA \cite{Rodrigo:2001kf,Czyz:2002np,Czyz:PH03}, EKHARA\cite{CN03} ),
 were and are being
 developed. The analysis presented
 in this paper is based on the results obtained by means of the program
 PHOKHARA 3.0 \cite{Czyz:PH03} (For further extensive discussions
 of various aspects of the radiative return method not covered by this
 article see [26-33],while for related discussion of the scan method
 look \cite{Gluza:rad,Gluza:eta}).
%\cite{Kuhn:2001,Rodrigo:2001jr,Rodrigo:2001cc,Kuhn:2002xg,Rodrigo:2002hk,Gluza:rad}.).
 
 Further complication arises as the photons and/or lepton pairs are 
 emitted also from final state charged hadrons and, 
 as that emission is model dependent,
 it has to be tested experimentally.
 That problem is discussed extensively in the next section.
\begin{figure}[ht]
\begin{center}
\epsfig{file=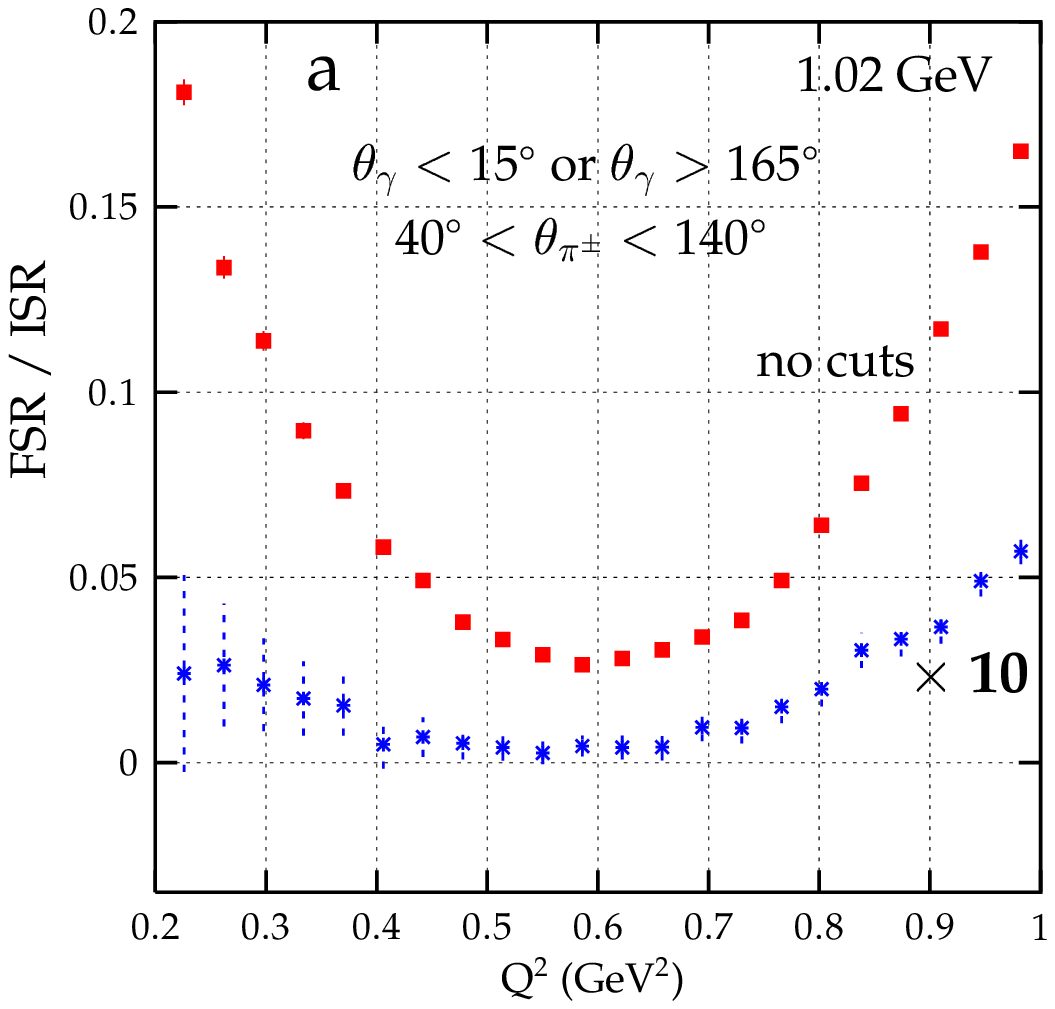,width=6.9cm,height=6cm}
\epsfig{file=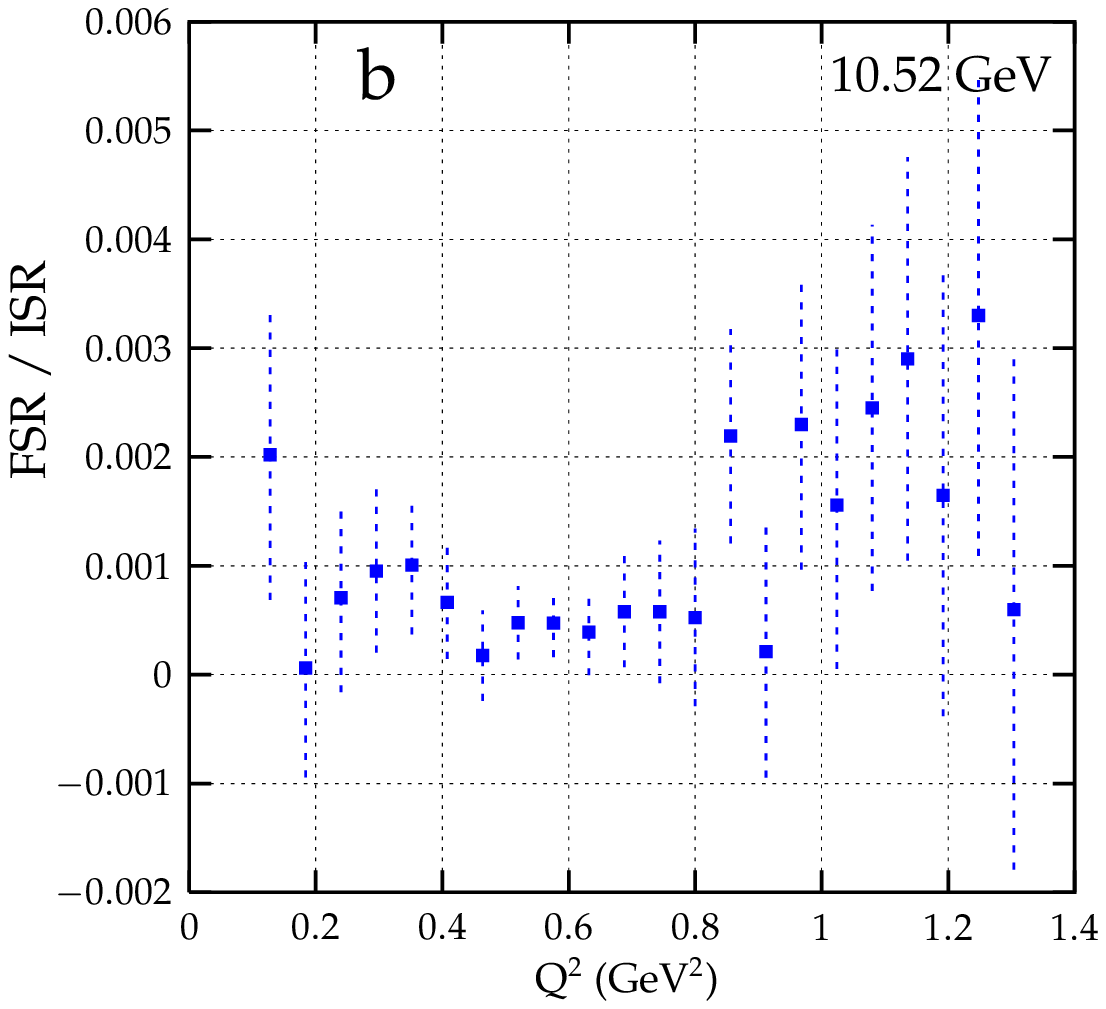,width=6.9cm,height=6cm}
\caption{Relative contribution of FSR with respect to ISR
 to the inclusive photon spectrum at \(\sqrt{s}~=\) 1.02 GeV
 (without and with cuts (multiplied by a factor 10)) (a)
  and  \(\sqrt{s}~=\) 10.52 GeV (b).}
\label{fig1}
\end{center}
\end{figure}

\section{FSR at LO and NLO}

\begin{figure}[htb]
\begin{center}
\epsfig{file=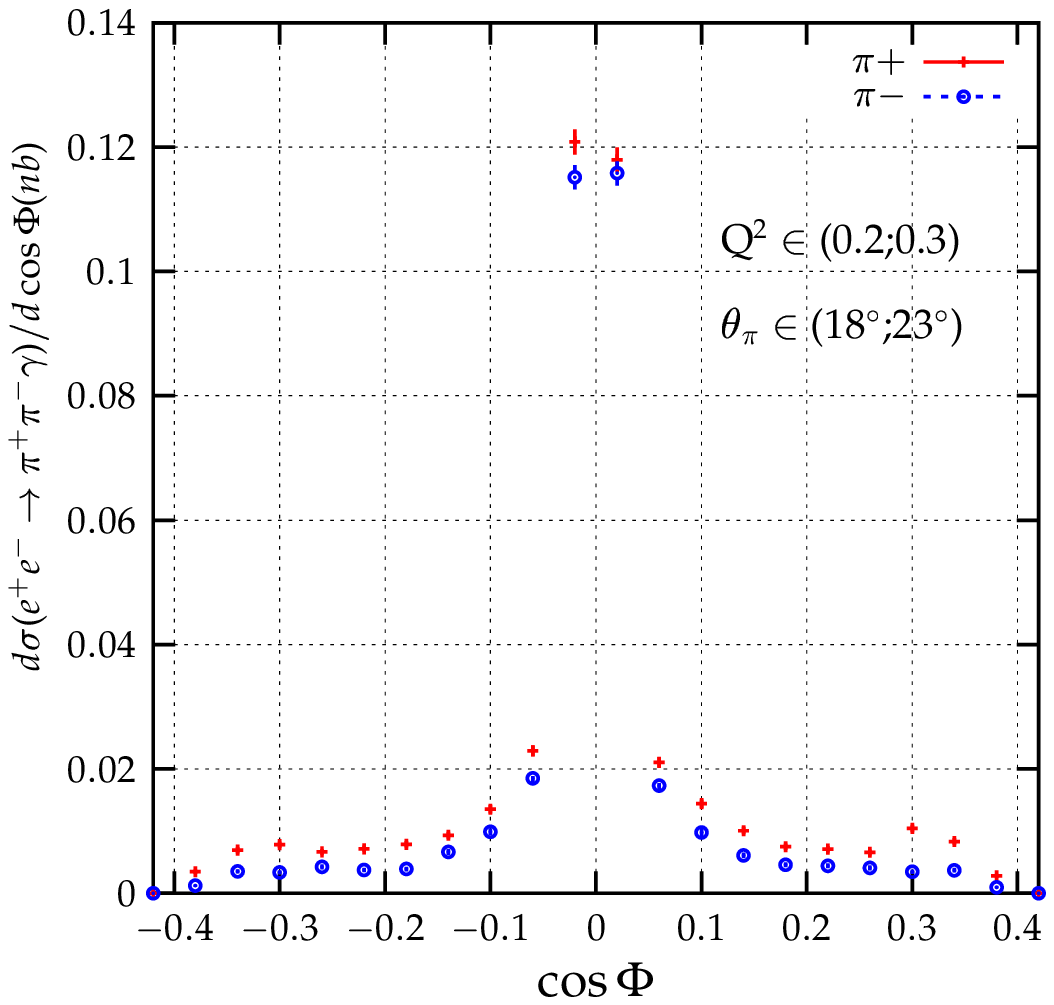,width=6.9cm,height=6cm}
\epsfig{file=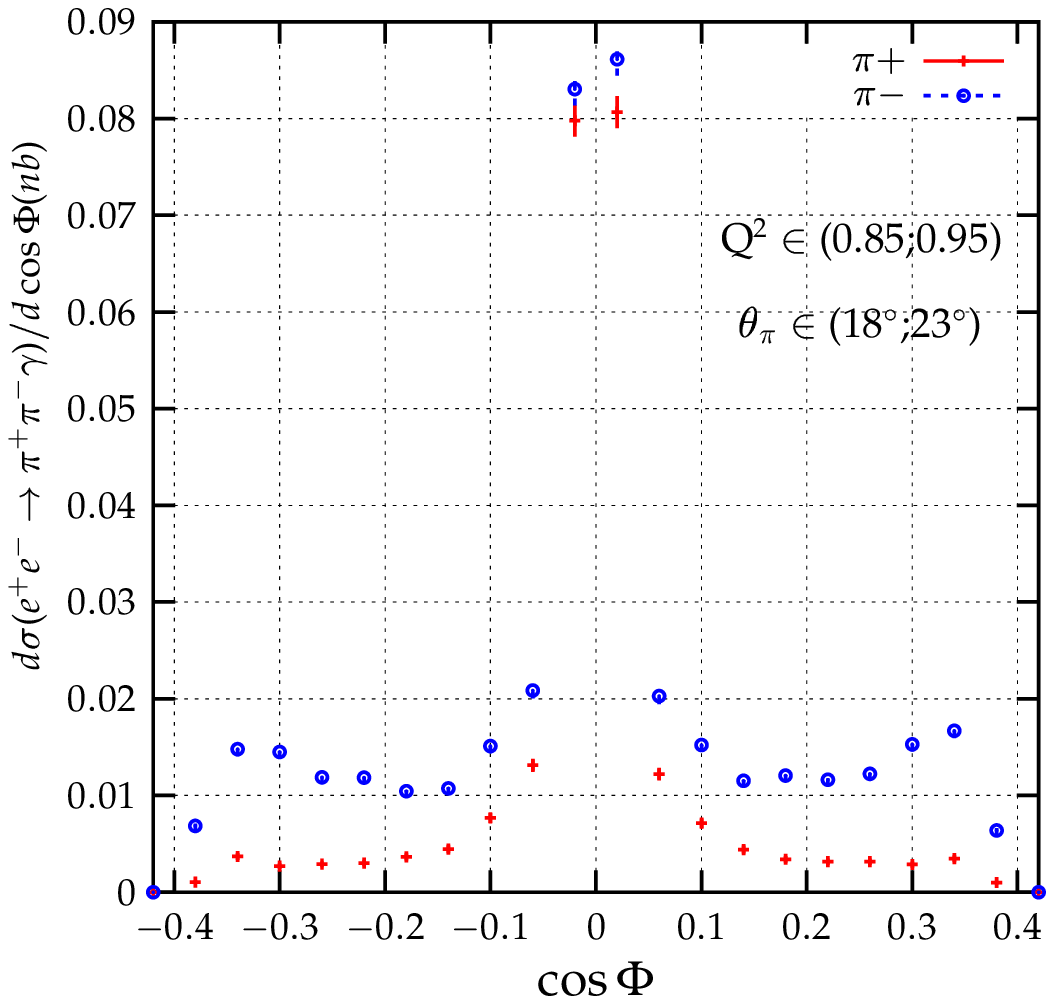,width=6.9cm,height=6cm}
\caption{Charge asymmetric distributions of the pions in $\Phi$ (the angle
 between normal to the production plain and the direction 
 of the initial positron) for different values of $Q^2$ 
 and fixed pion polar angles.}
\label{fig5}
\end{center}
\end{figure}

\begin{figure}[htb]
\begin{center}
\epsfig{file=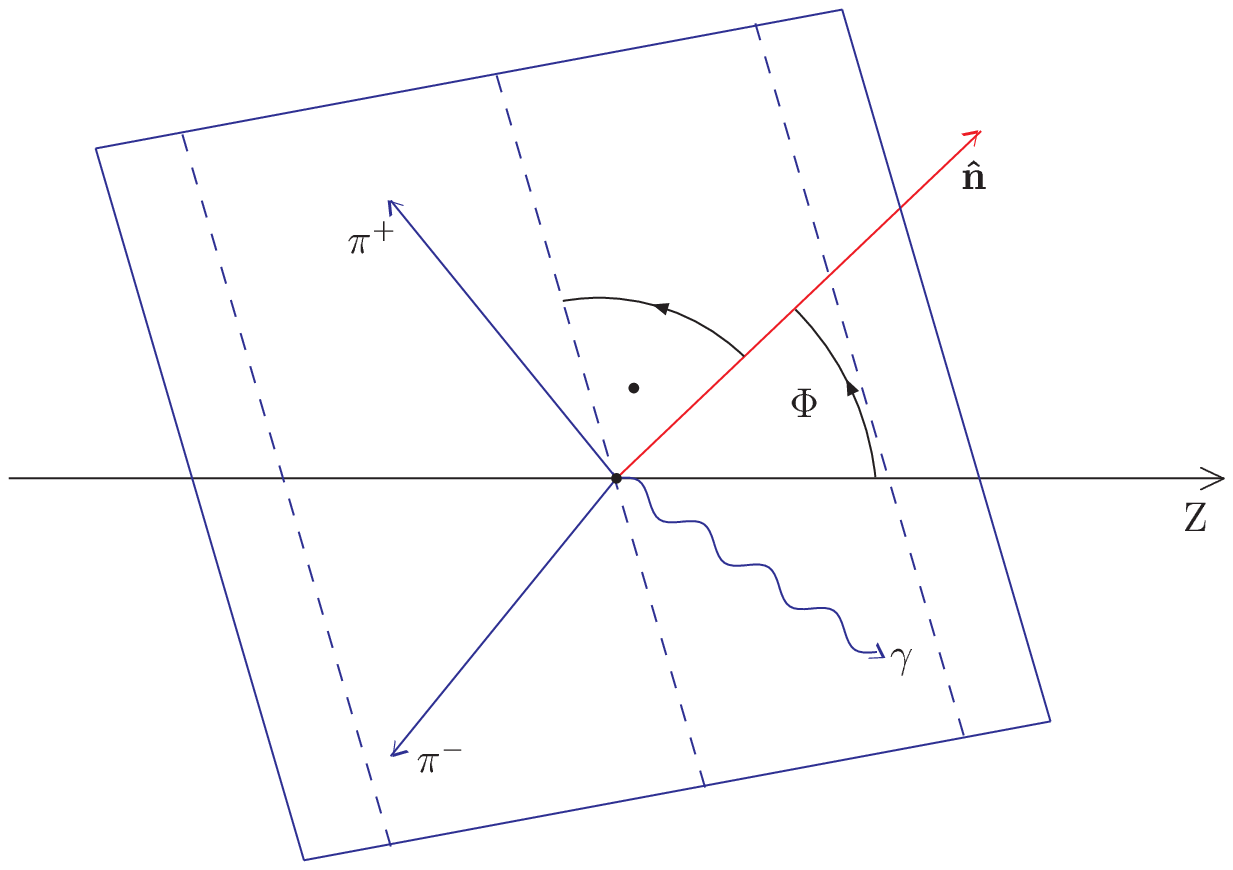,width=6.9cm,height=5.5cm}
\caption{Angle $\Phi$ defined as the angle
 between normal to the production plain and the direction 
 of the initial positron.}
\label{fig4}
\end{center}
\end{figure}

 At the leading order (LO) FSR contributions 
 to the process $e^+e^-\to\pi^+\pi^-\gamma$
 can be easily controlled by suitable cuts at $\Phi$--factories
 as shown in Fig.\ref{fig1}a
 and are completely negligible at B--factories (see Fig.\ref{fig1}b).
  The remaining FSR contribution at $\Phi$--factory, which is
  less than 1\%, can be subtracted from the data, relying on a MC generator,
  and the procedure of the
 hadronic cross section extraction described in the previous section can
 be used after that subtraction.
  The FSR contribution is however model dependent
 and one needs an independent experimental check on the accuracy of the
 model used.
 A simple observation that
 the interference of  ISR, which leads to a C-odd 
(C stands for charge conjugation) configuration of $\pi^+\pi^-$ pair,
  with the FSR amplitude, corresponding to C-even
 $\pi^+\pi^-$ configuration,
 vanishes if a charge symmetric event selection is used, but  
 it gives rise to charge asymmetries and charge induced 
 forward--backward asymmetries, is crucial for that tests.
 By relaxing the cuts and measuring various charge asymmetric
 distributions, extensive tests of FSR models are possible, and
 as the actual contribution of FSR to the radiative
 return cross section,  
 is of the order of 1\%, a modest 10\% accuracy of the model will lead to
 an error of 0.1\%, sufficient for any high precision measurement.
 Some of the tests of the model used in EVA and PHOKHARA for FSR
 (point-like pions and scalar QED (sQED)), proposed in \cite{Binner:1999bt},
  were already done by KLOE \cite{Aloisio:2001xq}, where it was
  shown that the charge asymmetry
\begin{equation}
 A(\theta) = \frac{N^{\pi^+}(\theta)-N^{\pi^-}(\theta)}
{N^{\pi^+}(\theta)+N^{\pi^-}(\theta)} \ \ ,
 \label{asym}
\end{equation}
 agrees well with the EVA MC \cite{Binner:1999bt}.
 However additional tests are needed to assure the accuracy
 of the model at the required level
 and comparisons of various charge asymmetric distributions
 between experimental data and MC are indispensable, especially
 when a measurement will use configurations with higher FSR contribution,
 necessary to cover the region of low $Q^2$.
 If only pions four-momenta are measured, as done in the KLOE experiment
 at the moment (see  \cite{Achim:radcor02,KLOE:2003}), one arrives at
  distributions as shown in Fig.\ref{fig5}. With
 500 pb$^{-1}$, collected till now by
 KLOE, the 0.1 nb/bin in the plot
 corresponds to 2000 events per bin. Thus that kind of measurement is 
 feasible and tests can be done with the required precision, provided
 systematic errors are small enough. The nontrivial $Q^2$ and polar angle
 dependence of $\Phi$--distributions 
 (definition of angle $\Phi$ is shown in Fig. \ref{fig4}) provides  profound
 cross checks of the tested model.

\begin{figure}[ht]
\begin{center}
\hskip 2 cm\epsfig{file=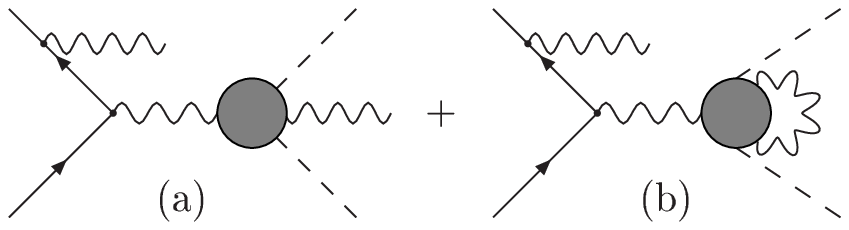,width=7.5cm,height=3.5cm} 
\caption{NLO contributions to the reaction
$e^+e^-\to\pi^+\pi^-\gamma$ from real (both soft and hard)
 FSR emission
  (a) and virtual corrections
 to the $\pi^+\pi^-\gamma^*$ vertex  (b).}
\label{fig9}
\end{center}
\end{figure}

\begin{figure}[ht]
\begin{center}
\epsfig{file=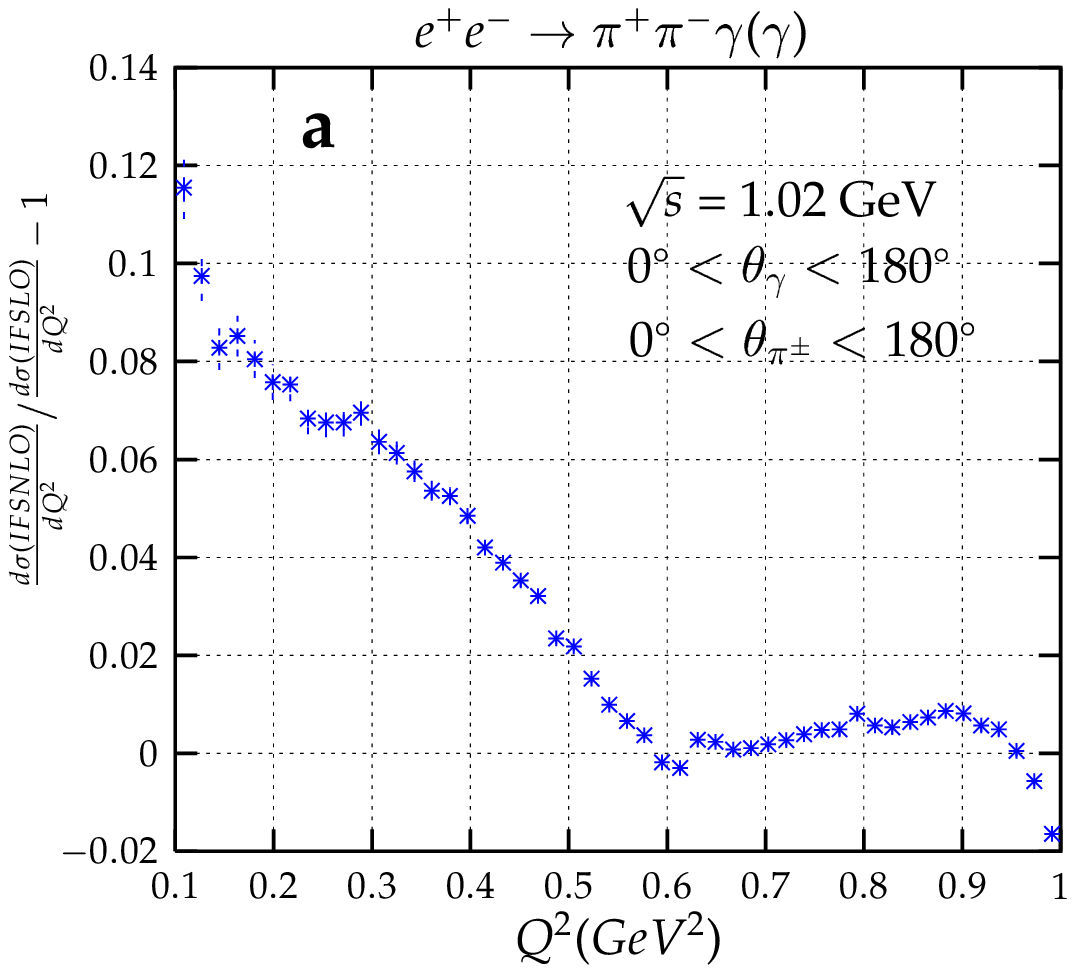,width=7.2cm,height=5.5cm}
\epsfig{file=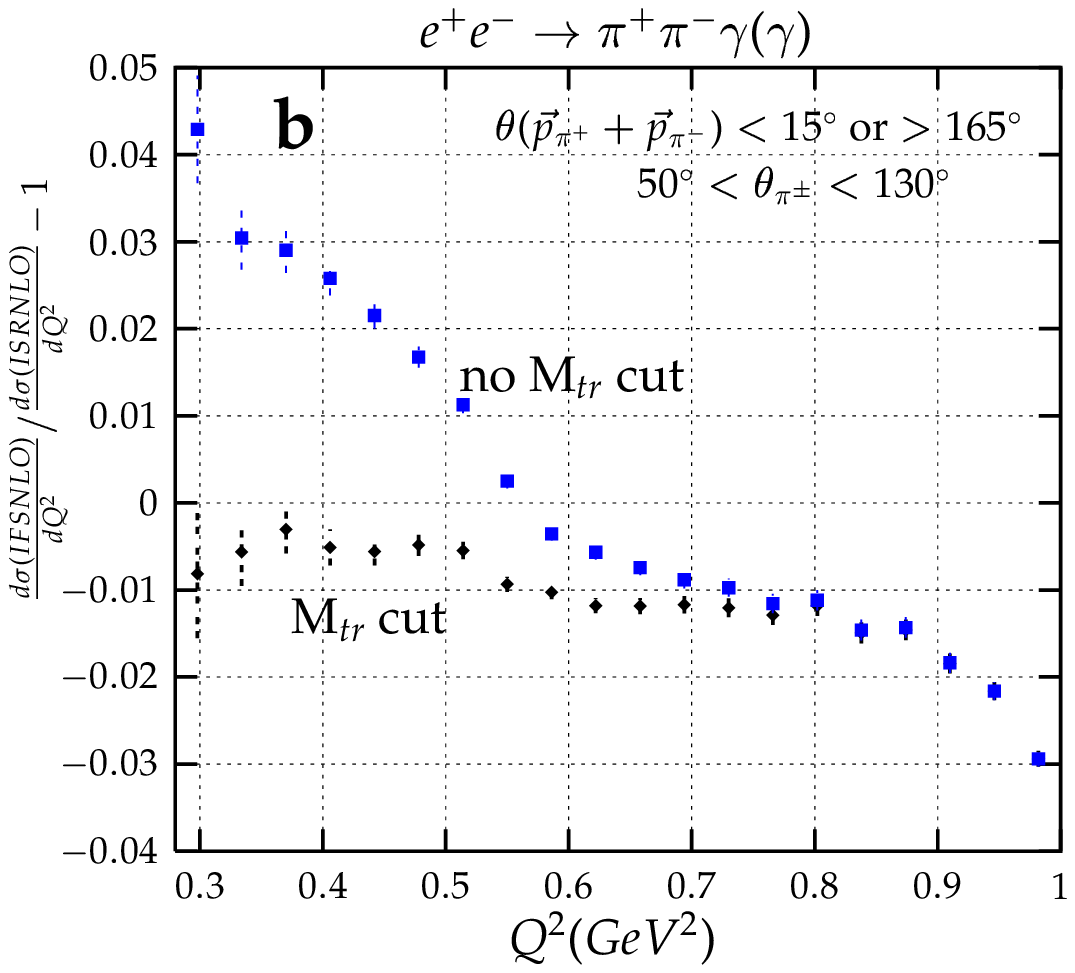,width=7.2cm,height=5.5cm}
\caption{Comparison of the \(Q^2\) differential cross sections
for \(\sqrt{s} = 1.02\)~ GeV:
IFSNLO contains the complete NLO contribution, while IFSLO has 
FSR at LO only. 
The pion and photon(s) angles are not
restricted in (a). In (b) cuts are imposed on
 the missing momentum direction and the track mass (see text for description).}
\label{fig10}
\end{center}
\end{figure}

\begin{figure}[ht]
\begin{center}
\epsfig{file=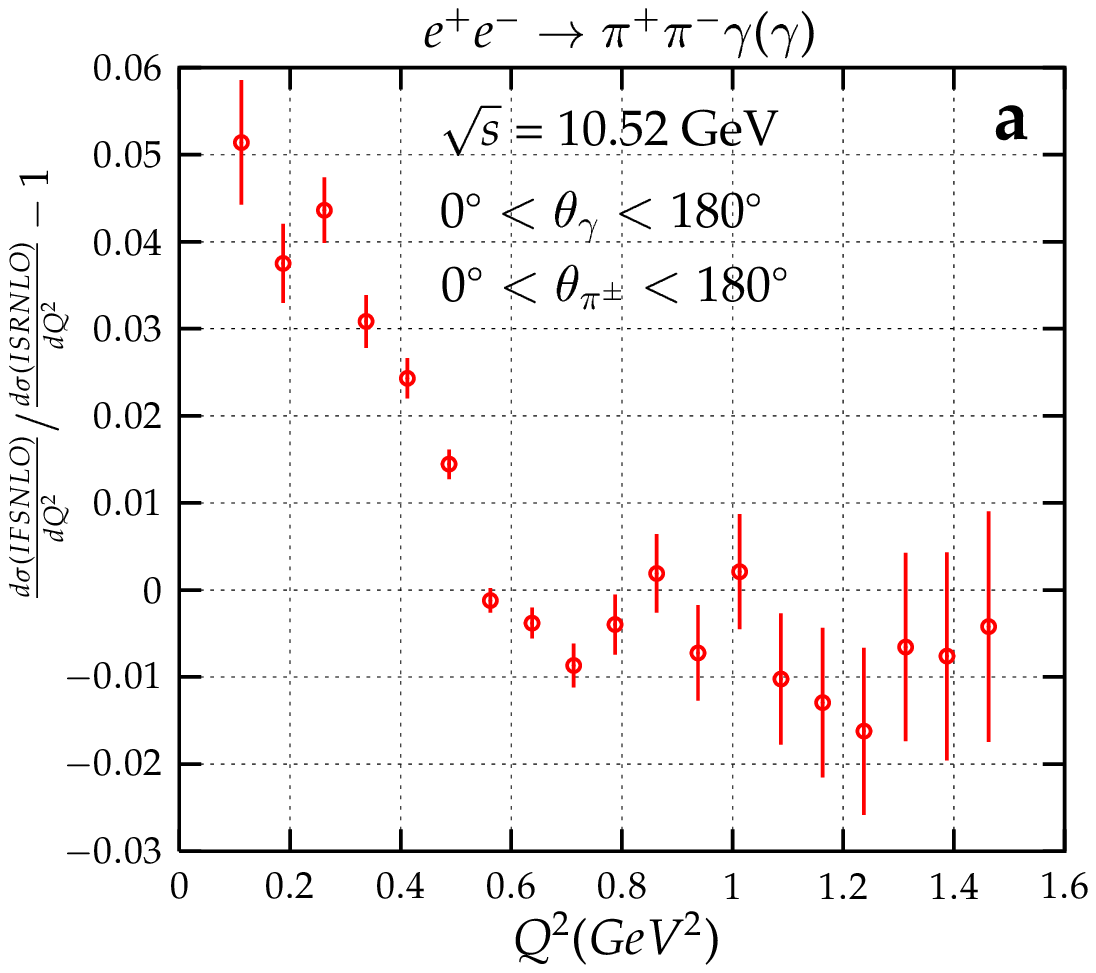,width=7.2cm,height=5.5cm}
\epsfig{file=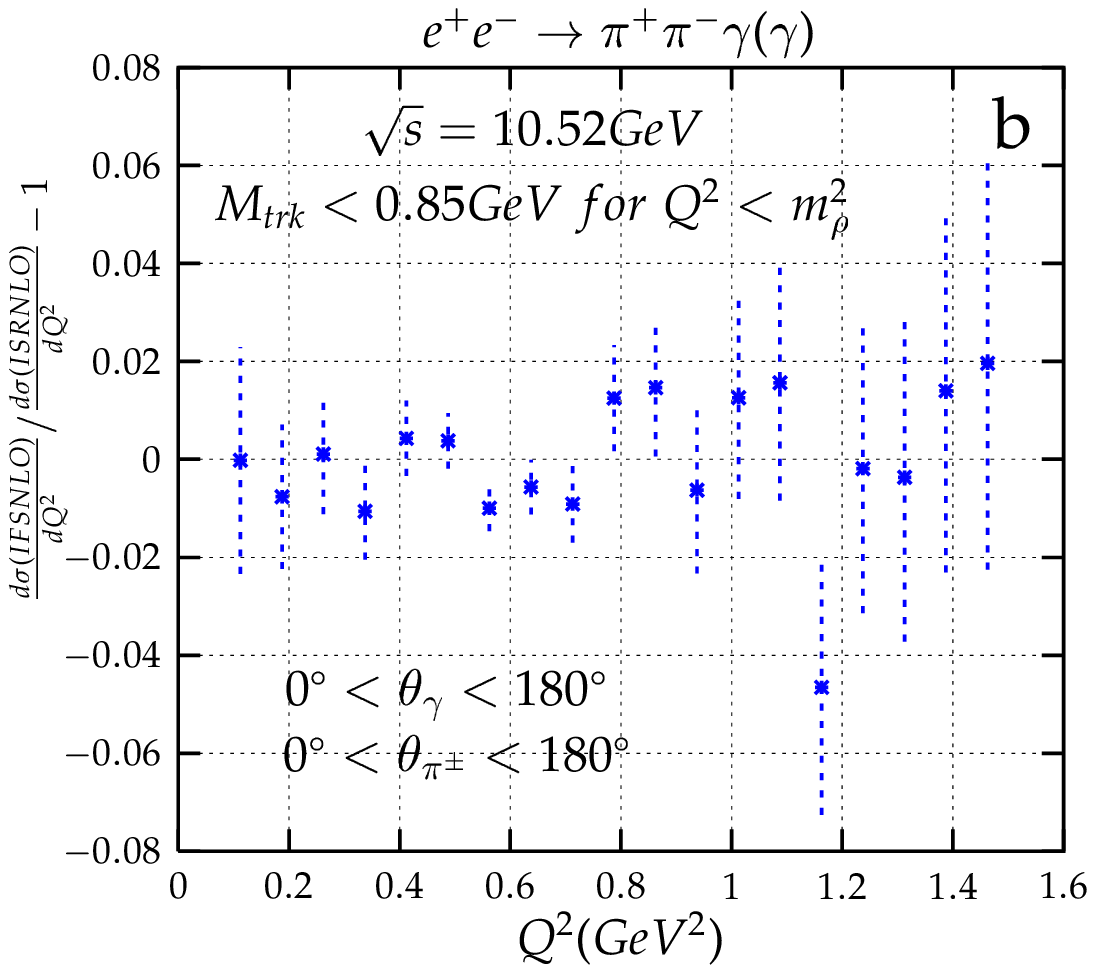,width=7.2cm,height=5.5cm}
\caption{Comparison of the \(Q^2\) differential cross sections
for \(\sqrt{s} = 10.52\)~ GeV:
IFSNLO contains the complete NLO contribution, while IFSLO has 
FSR at LO only. 
The pion and photon(s) angles are not
restricted in (a). In (b) cuts are imposed on
  the track mass for \(Q^2<m_\rho^2\) .}
\label{fig11}
\end{center}
\end{figure}

 At the next-to-leading order (NLO) the relevant diagrams contributing
 to the studied process are shown schematically in Fig.\ref{fig9}.
 They were implemented in the PHOKHARA event generator version 3.0 
\cite{Czyz:PH03}.
 From the analysis
 of the corresponding corrections to the $e^+e^-\to\pi^+\pi^-$ process
 \cite{Schwinger:ix} in the framework of sQED, this contribution 
 is expected to be of the order of 1\%.
 However, the emission of the initial photon reduces
 the invariant mass of the $\pi^+\pi^-$ (or $\pi^+\pi^-\gamma$) system
 to the $\rho$ mass with high probability due 
 to the peak of the pion form factor at the $\rho$ mass.
 As a result, 
 this contribution is strongly enhanced in the region of invariant mass of
 the $\pi^+\pi^-$ system below the $\rho$ resonance,
  as shown in Fig.\ref{fig10}a for KLOE energy
 and in Fig.\ref{fig11}a for B--factory energy.
 Suitably chosen cuts can be applied to suppress these NLO FSR 
 contributions. In Fig.\ref{fig10}b one can see that the standard KLOE
 cuts \cite{KLOE:2003}, which consist of the cuts
 on pion angles, the missing momentum angle and the track mass ($M_{tr}$),
 keep the NLO FSR contribution below 2\% with respect to the ISR cross section
 in the whole interesting region of the two--pion invariant mass.
 Similarly at B--factories, applying the track mass cut 
 only for events with $Q^2<m_\rho^2$, 
 the NLO FSR contribution is kept at a negligible level (Fig.\ref{fig11}b).
 The radiative corrections to $e^+e^-$ vertex with photon emitted from
 final state, which are not included in version 3.0 and are included
 in PHOKHARA 4.0 \cite{PHOKHARA_4.0} , might be as big
 as 2\%, if no cuts are applied, but are well below 0.1\% for the standard
 KLOE cuts.

\begin{figure}[ht]
\begin{center}
\epsfig{file=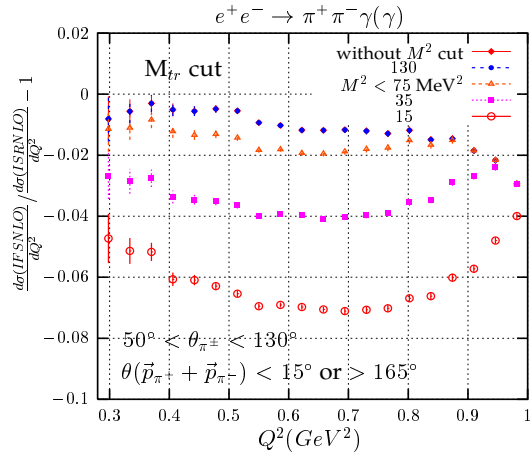,width=7.2cm,height=6cm}
\caption{Dependence of the relative IFSNLO contribution
 on the cut on missing invariant mass \(M^2\).}
\label{fig12}
\end{center}
\end{figure}

 Again, as in the case of LO FSR contributions, the main problem consists
 in the model dependence of FSR. Till now only few tests were performed
 to verify the model for FSR. However, if one aims at a measurement
 of the accuracy below 1\% such tests become indispensable.
 In the present KLOE experimental setup, where only four momenta of the 
 pions are measured, a possibility to test the hard part of the
 NLO FSR 
 contribution is to look at the dependence of the cross section
 on the missing invariant mass. Completely different effect of
 the cut on missing invariant 
 mass on ISR (ISRNLO) and FSR at NLO, as shown in Fig.\ref{fig12},
 provides a powerful tool for testing the hard part of the 
 IFSNLO contributions. A measurement of this few percent effect,
 depending on 
 the two--pion invariant mass $Q^2$ is within reach 
 of the KLOE experiment and a 
 detailed discussion of possible tests can be found
 in \cite{Czyz:PH03,CG03}.

\section{Conclusions}

 Huge event rates for $R(Q^2)$ measurements via radiative return
 method are, and will be  available in near future at $\Phi$-- and B--
 factories. They cover large range of $Q^2$ allowing for significant
 reduction of theory errors on muon anomalous magnetic moment and the running
 electromagnetic coupling. Experimental studies of the model dependence
 of the FSR contributions will allow for reduction of the error coming
 from that contribution to a negligible level.

 {\bf Acknowledgements:}
 This contribution is based on work performed in collaborations with 
 J.H.~K\"uhn and G.~Rodrigo. Henryk Czy\.z is grateful to 
 organisers for invitation, partial support and stimulating atmosphere of
 the workshop.

%%%%%%%%%%%%%%%%%%%%%%%%%%%%%%%%%%%%%%%%%%%%%%%%%%%%%%%%%%%%%%%%

\end{document}